\documentclass[dvips]{article}
\usepackage{supertabular,lscape,epsfig}
\usepackage{amssymb}
\usepackage{amsmath}

\DeclareSymbolFont{ppa}{OT1}{ppl}{m}{it}
\DeclareMathSymbol{\vv}{\mathalpha}{ppa}{'166}

\begin{document}

\newcommand{\dd}{\,{\rm d}}
\newcommand{\ie}{{\it i.e.},\,}
\newcommand{\etal}{{\it et al.\ }}
\newcommand{\eg}{{\it e.g.},\,}
\newcommand{\cf}{{\it cf.\ }}
\newcommand{\vs}{{\it vs.\ }}
\newcommand{\zdot}{\makebox[0pt][l]{.}}
\newcommand{\up}[1]{\ifmmode^{\rm #1}\else$^{\rm #1}$\fi}
\newcommand{\dn}[1]{\ifmmode_{\rm #1}\else$_{\rm #1}$\fi}
\newcommand{\upd}{\up{d}}
\newcommand{\uph}{\up{h}}
\newcommand{\upm}{\up{m}}
\newcommand{\ups}{\up{s}}
\newcommand{\arcd}{\ifmmode^{\circ}\else$^{\circ}$\fi}
\newcommand{\arcm}{\ifmmode{'}\else$'$\fi}
\newcommand{\arcs}{\ifmmode{''}\else$''$\fi}
\newcommand{\MS}{{\rm M}\ifmmode_{\odot}\else$_{\odot}$\fi}
\newcommand{\RS}{{\rm R}\ifmmode_{\odot}\else$_{\odot}$\fi}
\newcommand{\LS}{{\rm L}\ifmmode_{\odot}\else$_{\odot}$\fi}

\newcommand{\Abstract}[2]{{\footnotesize\begin{center}ABSTRACT\end{center}
\vspace{1mm}\par#1\par
\noindent
{~}{\it #2}}}

\newcommand{\TabCap}[2]{\begin{center}\parbox[t]{#1}{\begin{center}
  \small {\spaceskip 2pt plus 1pt minus 1pt T a b l e}
  \refstepcounter{table}\thetable \\[2mm]
  \footnotesize #2 \end{center}}\end{center}}

\newcommand{\TableSep}[2]{\begin{table}[p]\vspace{#1}
\TabCap{#2}\end{table}}

\newcommand{\FigCap}[1]{\footnotesize\par\noindent Fig.\  %
  \refstepcounter{figure}\thefigure. #1\par}

\newcommand{\TableFont}{\footnotesize}
\newcommand{\TableFontIt}{\ttit}
\newcommand{\SetTableFont}[1]{\renewcommand{\TableFont}{#1}}

\newcommand{\MakeTable}[4]{\begin{table}[htb]\TabCap{#2}{#3}
  \begin{center} \TableFont \begin{tabular}{#1} #4 
  \end{tabular}\end{center}\end{table}}

\newcommand{\MakeTableSep}[4]{\begin{table}[p]\TabCap{#2}{#3}
  \begin{center} \TableFont \begin{tabular}{#1} #4 
  \end{tabular}\end{center}\end{table}}

\newenvironment{references}%
{
\footnotesize \frenchspacing
\renewcommand{\thesection}{}
\renewcommand{\in}{{\rm in }}
\renewcommand{\AA}{Astron.\ Astrophys.}
\newcommand{\AAS}{Astron.~Astrophys.~Suppl.~Ser.}
\newcommand{\ApJ}{Astrophys.\ J.}
\newcommand{\ApJS}{Astrophys.\ J.~Suppl.~Ser.}
\newcommand{\ApJL}{Astrophys.\ J.~Letters}
\newcommand{\AJ}{Astron.\ J.}
\newcommand{\IBVS}{IBVS}
\newcommand{\PASP}{P.A.S.P.}
\newcommand{\Acta}{Acta Astron.}
\newcommand{\MNRAS}{MNRAS}
\renewcommand{\and}{{\rm and }}
\section{{\rm REFERENCES}}
\sloppy \hyphenpenalty10000
\begin{list}{}{\leftmargin1cm\listparindent-1cm
\itemindent\listparindent\parsep0pt\itemsep0pt}}%
{\end{list}\vspace{2mm}}

\def\TYLDA{~}
\newlength{\DW}
\settowidth{\DW}{0}
\newcommand{\dw}{\hspace{\DW}}

\newcommand{\refitem}[5]{\item[]{#1} #2%
\def\REFARG{#3}\ifx\REFARG\TYLDA\else, {\it#3}\fi
\def\REFARG{#4}\ifx\REFARG\TYLDA\else, {\bf#4}\fi
\def\REFARG{#5}\ifx\REFARG\TYLDA\else, {#5}\fi.}

\newcommand{\Section}[1]{\section{#1}}
\newcommand{\Subsection}[1]{\subsection{#1}}
\newcommand{\Acknow}[1]{\par\vspace{5mm}{\bf Acknowledgements.} #1}
\pagestyle{myheadings}

\newcommand{\xrule}{\rule{0pt}{2.5ex}}
\newcommand{\xxrule}{\rule[-1.8ex]{0pt}{4.5ex}}
\def\thefootnote{\fnsymbol{footnote}}

\newcommand{\TabCapp}[2]{\begin{center}\parbox[t]{#1}{\centerline{
  \small {\spaceskip 2pt plus 1pt minus 1pt T a b l e}
  \refstepcounter{table}\thetable}
  \vskip2mm
  \centerline{\footnotesize #2}}
  \vskip3mm
\end{center}}

\newcommand{\MakeTableSepp}[4]{\begin{table}[p]\TabCapp{#2}{#3}
  \begin{center} \TableFont \begin{tabular}{#1} #4 
  \end{tabular}\end{center}\end{table}}

\newfont{\bb}{ptmbi8t at 12pt}
\newfont{\bbb}{cmbxti10 at 12pt}
\newfont{\bbbb}{cmbxti10 at 9pt}
\newcommand{\uprule}{\rule{0pt}{2.5ex}}
\newcommand{\douprule}{\rule[-2ex]{0pt}{4.5ex}}
\newcommand{\dorule}{\rule[-2ex]{0pt}{2ex}}
\def\thefootnote{\fnsymbol{footnote}}
\begin{center}
{\Large\bf A Simple Method of Correcting Magnitudes for the Errors
Introduced by Atmospheric Refraction}

\vskip1cm

{\bf A.~~K~r~u~s~z~e~w~s~k~i~~~and~~~I.~~S~e~m~e~n~i~u~k}

\vskip3mm

{Warsaw University
Observatory, Al.~Ujazdowskie~4,~00-478~Warszawa, Poland\\
e-mail: (ak,is)@astrouw.edu.pl}
\end{center}
\vspace*{7mm}
\Abstract{We show that the errors due to atmospheric refraction are present in 
the magnitudes determined with the Difference Images Analysis method. In case 
of single, unblended stars the size of the effect agrees with the theoretical 
prediction. But when the blending is strong, what is quite common in a dense 
field, then the effect of atmospheric refraction can be strongly amplified to 
the extent that some cases of apparently variable stars with largest 
amplitudes of variations are solely due to refraction. We present a simple 
method of correcting for this kind of errors.}{}

\vskip10mm

It is generally known, that atmospheric refraction is a source of serious 
problems when one is doing astrometry (Evans and Irvin 1995 and references 
therein). The refraction shifts positions of observed stars toward the zenith, 
and as it is color dependent the resulting shifts of stars of different colors 
are different. 

The photometry is much less affected by refraction. Usually, a measurement of 
stellar magnitude is performed by three parameter fit of the stellar image to 
a known point spread function. Beside the magnitude that is the main objective 
of the measurement, two angular coordinates are also determined. An 
accompanying determination of the image position absorbs the effect of 
atmospheric refraction. The image position is shifted but the resulting value 
of magnitude is unaffected. In fact it is almost unaffected. The refraction 
causes the point spread function to be elongated in the direction toward the 
zenith. When the field size is large then the zenith distance varies and the 
degree of the  PSF elongation is changing when moving across the field. This 
can be handled by using any position dependent point spread function technique 
what in the case of large fields is usually necessary for other reasons anyway. 
There could also be some tiny dependence of the point spread function itself 
on spectral energy distribution of measured objects. An unresolved binary 
composed of stars with much different spectral types should have the point 
spread function more elongated in the direction to the zenith compared to the 
normal case of a single star. But such an effect of point spread function 
distortion should be small, unless the zenith distance is very large, and we 
are not aware of any attempt of taking it into account. 

The determination of angular coordinates that accompany the measurement of 
magnitude is a must when we use a single picture of a particular field. The 
coordinates are unknown and they have to be determined. The situation is 
different when we are investigating stellar variability using large number 
of CCD frames. Then, it is tempting when measuring magnitudes to use the 
fixed values for positions and to determine only a single parameter -- the 
magnitude. In this way we obtain an improvement in accuracy, as we get rid of 
additional degrees of freedom, and eventually have only one degree of freedom 
instead of three. In such a case the catalog of objects with -- presumably -- 
accurate positions has to be prepared beforehand based on a stacked average 
of frames or by averaging positions resulting from measurements of several 
frames. The use of the fixed values for coordinates may be even enforced at 
most cases of application of the image subtraction techniques. 

However, when magnitudes are determined for many epochs using fixed 
coordinates, then the values of resulting magnitudes are affected by movements 
of the apparent images. The object may move because it has sufficiently large 
proper motion or trigonometric parallax. It may also move due to color 
dependent atmospheric refraction. Since the introduction of the Difference 
Image Analysis the role of the refraction in the magnitude determination 
process has been often discussed (Tomaney and Crotts 1996, Melchior \etal 
1999, Alcock \etal 1999, Eyer and Wo{\'z}niak 2001). 

The set of OGLE-III observations used for finding transit candidates (Udalski 
\etal 2002) can serve as an example of the image  subtraction photometry with 
enforced fixed values for coordinates. In this case all stars have been 
photometered on each frame because the effect that was looked for was of a 
very small amplitude. And that was possible only with the use of an input 
catalog of fixed positions. This OGLE-III set is also exceptional by attaining 
relatively high accuracy what is uncommon for a mass survey work. This high 
accuracy in such kind of data made it possible to look for tiny systematic 
errors. In the remainder of this paper we shall show, based on the OGLE-III 
data, how to remove the disturbing influence of the differential refraction 
effect. In our calculations we have used only a small fraction of this 
OGLE-III data set, namely 5300 stars from the first chip of BLG100 field. 

\begin{figure}[p]
\centerline{\includegraphics[width=12.7cm,height=17cm]{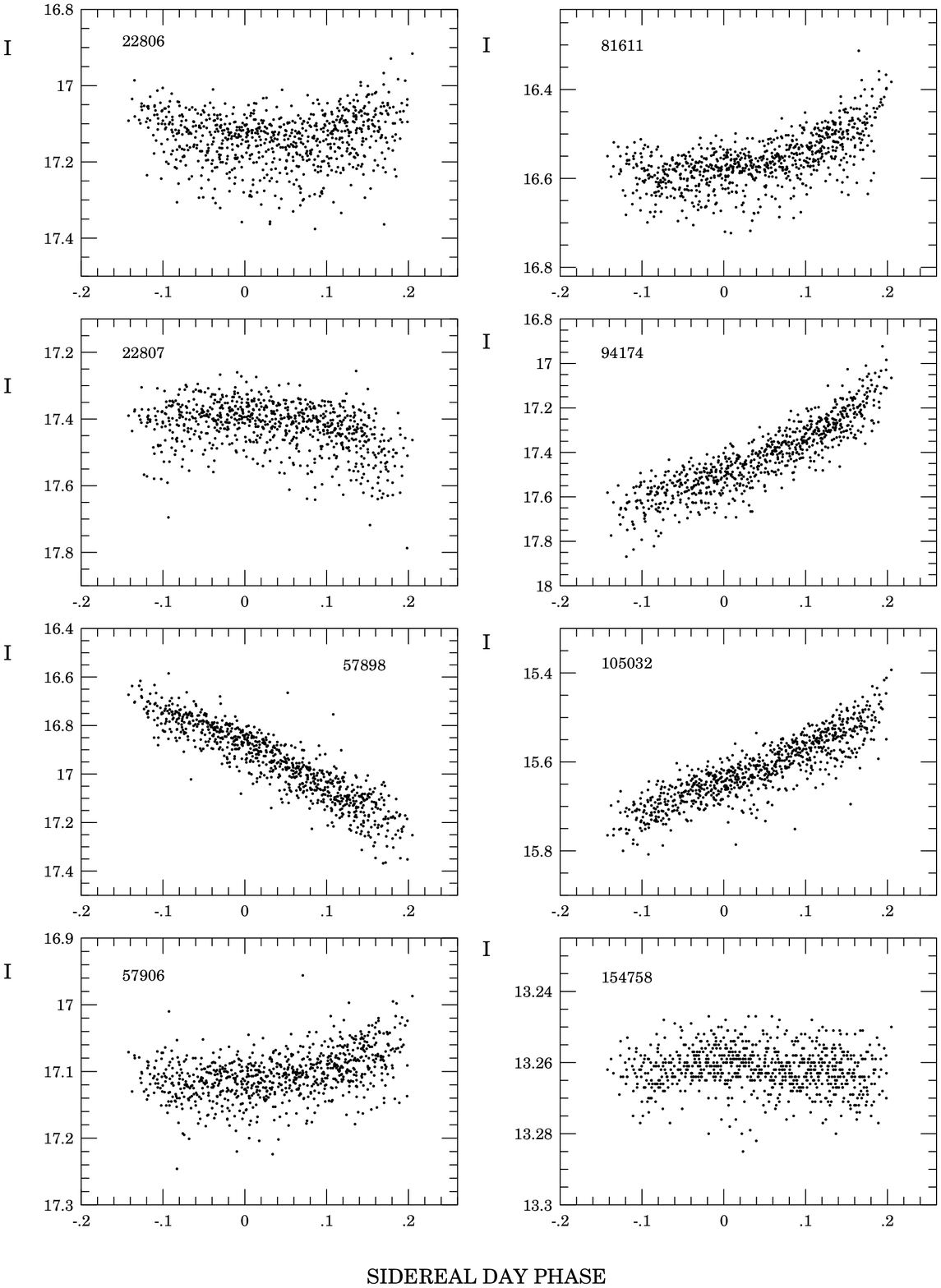}} 
\FigCap{The dependence of $I$ magnitude on the phase of sidereal day for a set 
of eight stars. The zero phase corresponds to the moment of the upper 
culmination of the field center.} 
\end{figure}

At first we have to estimate the expected size of the effect. We have done it 
in a rough way seeking only an order of magnitude estimate. As a starting 
point we selected 11 values of effective temperature spaced logarithmically in 
the range from 3000~K to 30~000~K. For each value of temperature a table of 
relative intensities was prepared as a function of wavelength according to the 
Planck function. This table was folded with sensitivity curves for the {\it V} 
and {\it I} filters and then folded again with the wavelength dependent 
refraction coefficient. As a result, we have obtained a set of refraction 
shifts in the filter {\it I} as a function of color index ${V-I}$. The outcome 
can be summarized as 0.01 arcsec difference for one magnitude difference in 
${V-I}$ color when observations are obtained at the zenith distance equal to 
45 degree. This, in case of the OGLE-III data, corresponds to 0.04 fraction of 
a pixel and in the case of sub-arcsecond seeing conditions may lead to a few 
millimagnitudes errors in measured magnitudes. 

Next we looked into the data in order to see how the observations are affected 
by the differential refraction effect. For this purpose we phased 
observations of individual stars with the sidereal day period. Well observed, 
unblended, constant stars with a color that is different from the average 
color should show an expected dependence of magnitude on sidereal day phase. 
Fig.~1 presents a set of well expressed dependences for eight stars. The zero 
phase corresponds to the moment of the upper culmination of the field center. 
Only the brightest among the stars in Fig.~1 can be considered unblended. This 
can apply to the star 154758 which shows the expected size of the atmospheric 
refraction effect. For all the other fainter stars, among those pictured in 
Fig.~1, it turns out that the effect of differential refraction is very 
strong, much stronger than expected, when there is a bright neighbor whose 
position is shifted by refraction. Small relative shifts in the neighbor 
position are changing the degree of blending, thus influencing the photometry 
of the fainter component. An example of blended objects is a pair of two first 
stars in Fig.~1 namely 22806 and 22807, whose positions are 10 pixels apart 
and there is another brighter star situated roughly in the middle of them. In 
the all remaining five cases there are brighter components closer than 4.5 
pixels or 1\zdot\arcs1.

The visual inspection of many plots like those presented in Fig.~1 led us to 
the conclusion that a sufficient description of the data can be accomplished 
with the help of a third degree polynomial 
$$m(h)=m_0+a_1h+a_2h^2+a_3h^3=m_0+r(h)\eqno(1)$$
where $h$ is an hour angle expressed in units of the sidereal day length,  
$m(h)$ is a stellar magnitude as a function of an hour angle, $m_0$ is the 
value of magnitude at the moment of the upper culmination, and the 
coefficients $a_1$, $a_2$, and $a_3$ are to be determined by the least squares 
fit to observations. 

\begin{figure}[p]
\centerline{\includegraphics[width=12.7cm,height=17cm]{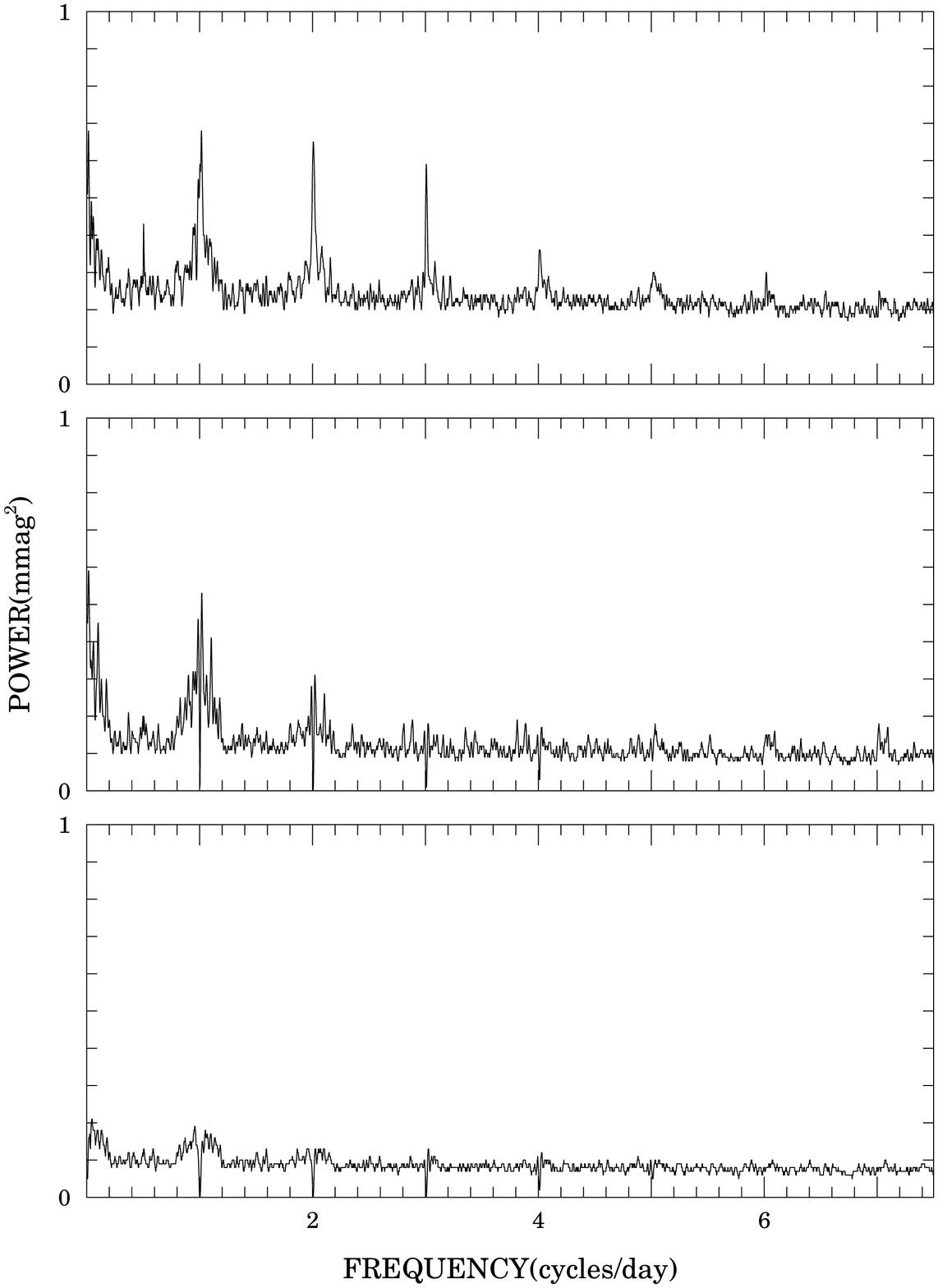}} 
\FigCap{The average periodogram for 200 bright, non-variable stars from the 
investigated field. The upper panel presents the periodogram, based on 
uncorrected data. The middle panel shows the average periodogram after 
correcting the observations of individual stars for refraction according to 
Eq.~(1), the lower panel presents the periodogram based on the observations 
corrected according to Eq.~(2).} 
\end{figure}

A signature of the differential refraction can be also traced in periodograms. 
We have calculated an average of periodograms for 200 bright, accurately 
observed, non-variable stars. As expected, this average shows peaks positioned 
at frequencies that are multiples of the sidereal day frequency. The average 
periodogram based on uncorrected data is plotted in the upper panel of Fig.~2. 
Next, from the magnitudes of each of these 200 stars we subtracted the fitted 
third order polynomial $r(h)$ defined in Eq.~(1), calculated periodograms for 
individual stars and then an average periodogram. This average  periodogram is 
plotted in the middle panel of Fig.~2. The peaks related to differential 
refraction disappear. The periodogram has zero values at frequencies exactly 
equal to multiple values of the sidereal day frequency. However, there is a 
raised level of noise around each of these zero values. The values of 
periodogram at frequencies close to zero are also higher than average. This is 
what is used to be called a red colored noise. Even seemingly constant stars 
may exhibit low amplitude oscillations with periods longer than 5 days. 
Another source of this excess of low frequency noise may be due to non 
negligible shifts connected with proper motions. In order to  limit this 
excess of low frequency noise we performed another fit to individual stars. 
This time beside the third degree polynomial with respect to the sidereal day 
phase we have included also a third degree polynomial with respect to time 
according to the formula 
$$m(h,t)=m_0+r(h)+b_1(t-t_0)+b_2(t-t_0)^2+b_3(t-t_0)^3\eqno(2)$$ 
which is Eq.~(1) extended by adding to it three new terms with up to the third 
power of the quantity $(t-t_0)$ where $t$ is mid-exposure time and $t_0$ 
is conveniently chosen to be an average of all mid-exposure times. 

The average of periodograms of the 200 stars calculated after the fit 
according to Eq.~(2) was subtracted from the raw observations is presented in 
the lower panel of Fig.~2. We can see that subtraction of the contribution 
from the differential refraction according to Eq.~(2) not only removes the 
peaks that originate from the differential refraction but also brings down the 
noise level and therefore makes it possible to detect variability of smaller 
amplitude. 

Usually any improvement has some price, sometimes in an unexpected form. Also 
in this case the procedure that causes an improvement for majority of objects 
may produce undesirable artifacts for some other objects. Of course it is 
commensurability between the sidereal day and a period of variability of a 
particular object that is a source of troubles. If the period of variability 
is in an exact resonance with the sidereal day then the situation is hopeless. 
Simply, we cannot detect such variability. In fact it is not much worrying. In 
the case of exact resonance, the observations cover only limited range of 
variability phases simply because they cover only limited range of sidereal 
day phases and therefore it is difficult to make a unique determination of the 
variability characteristics anyway. 

Let us consider a more common case when we are only close to the 
commensurability. We shall show that in such a case the damage made by the 
removal of differential refraction effect can be repaired. As an example of 
such object we take the star 143483. It is an eclipsing binary with the period 
of 0.66695 day so that we are very close to the 2:3 resonance. 

\begin{figure}[p]
\centerline{\includegraphics[width=12.7cm,height=16cm]{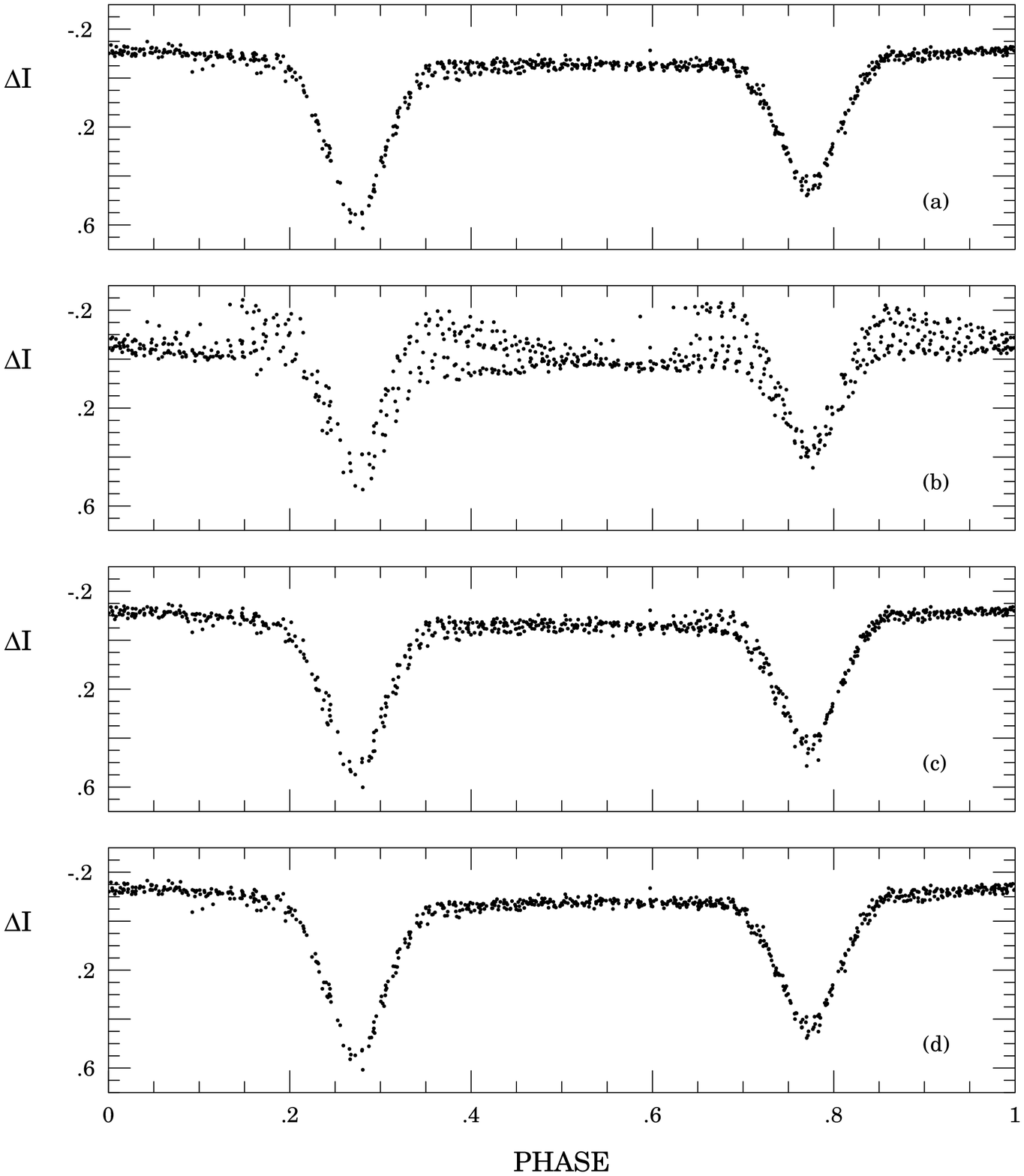}} 
\FigCap{Observations of eclipsing binary 143483. $\Delta I$ denotes the 
deviation of $I$ magnitude of the star from the mean value. The uppermost 
panel (a) shows the raw data uncorrected for refraction. Three remaining 
panels present the observations corrected for refraction. Panel (b) 
corresponds to the case when calculating the corrections only the polynomial 
describing refraction was fitted to the observations according to Eq.~(1). 
Panels (c) and (d) correspond to the situation, when calculating the 
refraction a sinusoid with the orbital period was fitted simultaneously to the 
observations according to Eq.~(3). Panel (c) is for the case when a sinusoid 
with the orbital period and its first overtone was fitted, and panel (d) for 
the case when first five even overtones of the orbital period were taken into 
account.} 
\end{figure} 

Fig.~3 shows in its uppermost panel (a) the eclipsing light curve using 
uncorrected observations. Panel (b) of Fig.~3 shows the same light curve but 
after the differential refraction correction was applied according to Eq.~(1). 
We can see the disastrous effect of the corrections. Two periods are so close 
to commensurability that the light variability due to eclipses produces 
spurious detection of the differential refraction effect. This happens when we 
force the fit to the formula that contains the terms $r(h)$ connected with the 
differential refraction only. It is different when we add to the fit also 
harmonic terms being functions of the phase $\theta$ of the eclipsing 
variations cycle and then subtract from the observations only $r(h)$. 
\setcounter{equation}{2}
\begin{eqnarray}
m(h,\theta) & = & m_0+r(h)+c_0\cos(2\pi\theta)+
            d_0\sin(2\pi\theta)\nonumber \\
            & &+\sum_{k=1}^n[c_k\cos(4k\pi\theta)+d_k\sin(4k\pi\theta)].
\end{eqnarray}

Panels (c) and (d) of Fig. 3 show the resulting light curves obtained with 
different numbers of harmonic expansions terms. The number of terms $n$ is 
equal to one in panel (c) and it is equal to five in the case of panel (d). 

We can conclude that when dealing with observations made with the use of the 
{\it I} filter the atmospheric refraction effect is small and comparable with 
observational errors in the case of bright, isolated, well observed stars. It
can be, however, dominating in the case of faint objects blended with a bright 
component. 

The atmospheric refraction is absent outside of the Earth atmosphere. 
Hopefully it can be also eliminated on the observation stage in case of 
ground-base observations. A device planned for VLT Survey Telescope and 
described in (Kuijken 2002) should do the job. We should remember however that 
we cannot get away from the disturbing effects of proper motions or parallaxes 
by using any intelligent device or by putting telescopes into space. 

\Acknow{The authors are greatly indebted to Bohdan Pa\-czy\'n\-ski and Andrzej 
Udalski for making possible to use the OGLE-III photometry. This work was 
partly supported by the KBN grant BST to the Warsaw University Observatory.}

\end{document}